\documentclass[12pt,thmsa]{article}
\usepackage{sw20lart}

\input tcilatex
\QQQ{Language}{
American English
}

\begin{document}

\author{Kaige Wang$^{1,2}$ and Shiyao Zhu$^2$ \\
1. CCAST (World Laboratory), P. O. Box 8730, Beijing 100080, \\
and Department of Physics, Beijing Normal University, Beijing 100875, China%
\thanks{%
Mailing address}\\
2. Department of Physics, Hong Kong Baptist University, Hong Kong, China}
\title{Storage States in Ultracold Collective Atoms}
\maketitle

\begin{abstract}
We present a complete theoretical description of atomic storage states in
the multimode framework by including spatial coherence in atomic collective
operators and atomic storage states. We show that atomic storage states are
Dicke states with the maximum cooperation number. In some limits, a set of
multimode atomic storage states has been established in correspondence with
multimode Fock states of the electromagnetic field. This gives better
understanding of both the quantum and coherent information of optical field
can be preserved and recovered in ultracold medium. In this treatment, we
discuss in detail both the adiabatic and dynamic transfer of quantum
information between the field and the ultracold medium.

PACS numbers: 42.50.Gy, 03.67.-a
\end{abstract}

\section{Introduction}

In the interaction between optical fields and atoms, a photon can be
absorbed by an atom and then the excited atom can re-emit a photon by either
spontaneous or stimulated emission. In this process the atom stores the
energy of the field and releases it back to the field. Recently, theoretical
and experimental studies have shown that both the quantum state and coherent
information of the field can be stored in an atomic medium \cite{liu}-\cite
{flei2}. A very recent experiment witnesses that a signal pulse of light can
be stored in an ultracold collective of atoms for up to a millisecond \cite
{liu}.

The basic scheme for storage of light information is carried out by
electromagnetically induced transparency (EIT) \cite{harris},\cite{arim}. $N$
three-level atoms with one upper level $|a\rangle $ and two lower levels $%
|b\rangle $\ and $|c\rangle $ interact resonantly with both the signal and
the control beams. The weak signal beam and the strong control beam drive
the atomic transitions $|a\rangle -|b\rangle $ and $|a\rangle -|c\rangle $,
respectively. Early investigations have shown that EIT permits the
propagation of the light signal through an otherwise opaque atomic medium
and that the group velocity of the signal pulse is greatly reduced \cite{hau}%
-\cite{bud}. In a recent experiment \cite{liu}, when the control beam is
turned off, the signal pulse is stopped and stored in atomic medium. This
effect can be understood in terms of dark states. Dark states are
combination states of the photon state and the atomic storage state. The
bosonic quasiparticles in the dark state are called polaritons. When the
strength of the control field is changed adiabatically, both the quantum
state and the coherent information transfer between the signal field and the
collective atoms \cite{flei},\cite{lukin}.

Due to the fact that dark states are eigenstates of the EIT interaction
Hamiltonian, the transfer of the quantum state between the field and the
atomic ensemble can be quasi-stationary in the adiabatic limit. Furthermore,
mapping and storage of quantum information of light in an atomic medium may
occur in a dynamical process without forming dark states. For example, Ref. 
\cite{pol} has shown that in a general Raman interaction, with a large
detuning to the intermediate level, the nonclassical features of the quantum
field can be mapped onto the coherence of the lower atomic doublet,
distributed over the atomic cloud. However, a very recent theoretical study
has shown that, in the EIT model, an adiabatic change of the control field
is not necessary, and even a fast switching of the control field can be used
in the writing and reading quantum information of the signal field \cite{mat}%
.

In all of these models, collective atoms play the role of quantum memory.
The mechanism of storage of the quantum field in a medium is based on
establishing atomic storage states (or atomic memory states) which record
all the information of the field. It is not surprising that photons can be
transferred to atomic excitations in transition interactions. Physically,
the question is how collective atoms record the coherence of an
electromagnetic field. In the early theoretical work \cite{dicke}, Dicke
studied the coherence and cooperation effects in atomic ensemble. He defined
collective atomic operators as the sum of all the individual atomic
operators, retaining the angular momentum properties. The Dicke states are
the eigenstates of the angular momentum operators. In Dicke's paper, he
considered two cases: the gas volumes have dimensions either smaller or
larger than the radiation wavelength. For the latter case, the spatial phase
distribution of the field has been included into the collective atomic
operators.

In the pioneering work contributed by M. Fleischhauer and M. D. Lukin \cite
{flei},\cite{lukin},\cite{lukin2},\cite{flei2}, the theory of atomic storage
states basically consists of a single-mode description. In this paper, we
present a complete description of atomic storage states in a multimode
framework. A preliminary version of the briefly analysis described in this
paper is given in Ref. \cite{wang}. Similarly, as in the second case of
Dicke's work, we incorporate spatial coherence into the collective atomic
lower and upper operators. We indicate that, in the multimode description,
atomic collective operators can behave as multimode bosonic operators, under
the conditions of low atomic excitation and appropriate radiation
wavelength; which is much larger than the average interval of atoms and less
than the propagation length of medium. Aside from the original definition of
Dicke states, we introduce atomic storage states with explicit expressions
by containing spatial coherence of the radiation field, and indicate that
they are Dicke states with the maximum cooperation number. The significant
advance is that, under the conditions shown above, a set of multimode atomic
storage states is established in correspondence with multimode Fock states
of the electromagnetic field. This gives a better understanding of how both
quantum and coherent information, of electromagnetic fields, can be
preserved in atomic media. A detailed theoretical description of multimode
dark states in the EIT model is discussed. Furthermore, parallel to the
''stationary polariton'' in EIT, we show the ''dynamic polariton'' formed in
coupled harmonic oscillators. This illustrates the mechanism for the dynamic
quantum transfer between field and macroscopic matter.

\section{Atomic collective operators with the bosonic commutation}

We consider $N$ ultracold collective atoms which are approximately
stationary at their positions. At very low temperature close to the critical
temperature for Bose-Einstein condensation\cite{liu}, the average kinetic
energy of atoms is greatly reduced. On the other hand, at low temperature,
atoms are densely packed within a wavelength of optical field. The free path
of an atom is much less than the wavelength, hence it confines the range of
atomic motion. What the ''still atoms'' means is that, in the characteristic
time of the system, the scale of motion for the centre-of-mass of the atoms
is much less than the wavelength of the optical electromagnetic field
involved. The two levels of atoms $|b\rangle $\ and $|c\rangle $ interact
with some optical field of wavevector $k$. We assume that $N$\ is a large
number and the largest proportion of the population of the atoms is in level 
$|b\rangle $ throughout the system evolution, so that the completeness
relation is given by 
\begin{equation}
N=\sum_{j=1}^N(|b_j\rangle \langle b_j|+|c_j\rangle \langle c_j|)\simeq
\sum_{j=1}^N|b_j\rangle \langle b_j|.  \label{1}
\end{equation}
It is not necessary that $|b\rangle $ is the ground state, for instance, in
the case of EIT the level $|c\rangle $ can be lower or equal to $|b\rangle $%
. For the sake of convenience, we call $|b\rangle $ the ''ground'' state and 
$|c\rangle $ the ''excited'' state.

In the interaction between field and atomic medium, the spatial coherence of
the field affects only the local atoms. In the approximation of ''still
atoms'', the $j$-th atom located at position $z_j$ suffers a local field
strength with a phase $\exp (ikz_j)$. For this reason, we define the lower
and the upper operators of the collective atoms as 
\begin{mathletters}
\label{2}
\begin{eqnarray}
\sigma _k &=&\frac 1{\sqrt{N}}\sum_{j=1}^N|b_j\rangle \langle c_j|\exp
(-ikz_j),  \label{2a} \\
\sigma _k^{\dagger } &=&\frac 1{\sqrt{N}}\sum_{j=1}^N|c_j\rangle \langle
b_j|\exp (ikz_j),  \label{2b}
\end{eqnarray}
where $k$ is the wavevector of the optical electromagnetic field interacting
with the transition $|b\rangle -|c\rangle $. We notice that this kind of
collective atomic operator, containing spatial coherence, was first
introduced by Dicke, who investigated the super-radiate effect in collective
atoms, in the case of medium dimensions larger than radiation wavelength 
\cite{dicke}. In order to avoid the difficulties of the occurrence of the
center-of-mass motion of atoms which may destroy the coherence, in Dicke's
paper, he assumed the molecules are so massive that their center-of-mass
coordinates will be then treated as time-independent parameter in equation.
Now, this assumption can be implemented in the development of the ultracold
technique.

For the purpose of controllable storage, the atomic transition $|b\rangle
-|c\rangle $ is usually a multi-photon process including signal and control
photons. Equation (\ref{2a}) should be replaced by 
\end{mathletters}
\begin{mathletters}
\label{2p}
\begin{eqnarray}
\sigma _k &=&\frac 1{\sqrt{N}}\sum_{j=1}^N|b_j\rangle \langle c_j|\exp
[-i(k-k_c)z_j],  \label{2pa} \\
\sigma _k &=&\frac 1{\sqrt{N}}\sum_{j=1}^N|b_j\rangle \langle c_j|\exp
[-i(k+k_c)z_j],  \label{2pb}
\end{eqnarray}
where $k$ and $k_c$ are respectively the wavevectors for the signal field
and the control field. Equation (\ref{2pa}) describes a Raman transition,
whereas Eq. (\ref{2pb}) describes a two-photon cascade transition.

The commutation relations for these atomic operators are written as 
\end{mathletters}
\begin{mathletters}
\label{3}
\begin{eqnarray}
\lbrack \sigma _k,\sigma _{k^{\prime }}] &=&[\sigma _k^{\dagger },\sigma
_{k^{\prime }}^{\dagger }]=0,  \label{3a} \\
\lbrack \sigma _k,\sigma _{k^{\prime }}^{\dagger }]
&=&(1/N)\sum_{j=1}^N(|b_j\rangle \langle b_j|-|c_j\rangle \langle c_j|)\exp
[-i(k-k^{\prime })z_j].  \label{3b}
\end{eqnarray}
The exact commutation of Eq. (\ref{3b}) for the same mode is readily
obtained 
\end{mathletters}
\begin{equation}
\lbrack \sigma _k,\sigma _k^{\dagger }]=(1/N)\sum_{j=1}^N(|b_j\rangle
\langle b_j|-|c_j\rangle \langle c_j|).  \label{3p}
\end{equation}
If $N$ is a very large number and most of the atomic population rests in
level $|b\rangle $ throughout evolution, by applying Eq. (\ref{1}), then Eq.
(\ref{3b}) is approximately reduced to 
\begin{equation}
\lbrack \sigma _k,\sigma _{k^{\prime }}^{\dagger }]\simeq
(1/N)\sum_{j=1}^N\exp [-i(k-k^{\prime })z_j].  \label{3q}
\end{equation}
Assuming that the atoms are in a string and the average interval of the
adjacent atoms is $d$, which is much less than the optical wavelength, i.e. $%
kd<<1$, one obtains 
\begin{equation}
\sum_{j=1}^N\exp [ikz_j]=\sum_{j=1}^N\exp [ik(j-1)d]=\frac{1-\exp [ikNd]}{%
1-\exp [ikd]}\approx N\frac{\exp [ikL]-1}{ikL},  \label{3r}
\end{equation}
where $L=Nd$ is the length of the atomic medium. However, this result is
also true for a volume of atomic gas which is considered as a continuous
medium 
\begin{equation}
\sum_{j=1}^N\exp [ikz_j]=\int_0^L\frac NL\exp [ikz]dz=N\frac{\exp [ikL]-1}{%
ikL}.  \label{3s}
\end{equation}
In the case that the length of the atomic medium is much larger than the
optical wavelength, we obtain 
\begin{equation}
\frac 1N\sum_{j=1}^N\exp [ikz_j]=\left\{ 
\begin{array}{c}
1\qquad \quad (k=0), \\ 
0\qquad (kL>>1).
\end{array}
\right.  \label{3t}
\end{equation}
By applying the above result to Eq. (\ref{3q}), one obtains the bosonic
commutation relation for the collective atomic operators 
\begin{equation}
\lbrack \sigma _k,\sigma _{k^{\prime }}^{\dagger }]\simeq \delta
_{kk^{\prime }},  \label{4}
\end{equation}
where we should assume $(k-k^{\prime })L>>1$, or, equivalently, $\lambda
-\lambda ^{\prime }>>\lambda ^2/(2\pi L)$. For the parameters used in the
experiment \cite{liu}, $L=339\mu m$ and $\lambda =589.6nm$, so that $\lambda
^2/(2\pi L)\approx 0.163nm$, Eq. (\ref{4}) is a good approximation for
distinguishable modes.

We summarize the conditions for the collective atomic operators satisfying
the multimode bosonic commutation as 
\begin{mathletters}
\label{c}
\begin{eqnarray}
N &\gg &n,  \label{ca} \\
\lambda ,L &\gg &d,  \label{cb} \\
\Delta \lambda /\lambda  &\gg &\lambda /L,  \label{cc}
\end{eqnarray}
where $n$ is the number of atomic excitations and $\Delta \lambda $ is the
mode interval. The low excitation limit (\ref{ca}) has already been shown in
the previous paper \cite{flei}. The other two conditions imposed on the
radiation wavelength, Eqs. (\ref{cb}) and (\ref{cc}), assure the atomic
ensemble containing and distinguishing the coherence, respectively. We will
see in the next section that the atomic collective operators behave
similarly to the creation and annihilation operators of the electromagnetic
field.

\section{Single-mode atomic storage states}

The ''ground-level'' state of the atoms can be compared with the ''vacuum''
state, symbolized in Ref. \cite{flei} as 
\end{mathletters}
\begin{equation}
|C^0\rangle \equiv |b_1b_2\cdots b_N\rangle .  \label{5}
\end{equation}
When the single-mode creation operators of the collective atoms are applied
to the ''vacuum''state, one obtains 
\begin{eqnarray}
(\sigma _k^{\dagger })^n|C^0\rangle &=&\frac 1{\sqrt{N^n}}\left(
\sum_{j=1}^N|c_j\rangle \langle b_j|\exp (ikz_j)\right) ^n|b_1b_2\cdots
b_N\rangle  \label{6} \\
&=&\frac 1{\sqrt{N^n}}\sum_{\{i_n\}}{}^{\prime }|c_{i_1}\cdots
c_{i_n}\rangle \langle b_{i_1}\cdots b_{i_n}|b_1b_2\cdots b_N\rangle \exp
[ik(z_{i_1}+\cdots +z_{i_n})]  \nonumber \\
&=&\frac 1{\sqrt{N^n}}\sum_{\{i_n\}}{}^{\prime }|b_1\cdots c_{i_1}\cdots
c_{i_n}\cdots b_N\rangle \exp [ik(z_{i_1}+\cdots +z_{i_n})]  \nonumber
\end{eqnarray}
where $\sum_{\{i_n\}}^{\prime }$ designates that, in the summation, any two
indices cannot be equal, because $(|c_j\rangle \langle b_j|)^2|b_j\rangle =0$%
. We note that some states in the summation of Eq. (\ref{6}), for which the
sequence in the index set $\{i_n\}$ is exchanged, are the same and should be
put together. For example, $(i_1=1,i_2=2,i_3,\cdots ,i_n)$ and $%
(i_1=2,i_2=1,i_3,\cdots ,i_n)$ represent the same state. For an ensemble $%
\{i_n\}$ of $n$\ elements, there are $n!$\ permutations which form the same
state. By eliminating these repeated terms in the summation, Eq. (\ref{6})
can be replaced by 
\begin{equation}
(\sigma _k^{\dagger })^n|C^0\rangle =\frac{n!}{\sqrt{N^n}}%
\sum_{\{i_n\}}{}^{\prime \prime }|b_1\cdots c_{i_1}\cdots c_{i_n}\cdots
b_N\rangle \exp [ik(z_{i_1}+\cdots +z_{i_n})],  \label{7}
\end{equation}
where $\sum_{\{i_n\}}^{\prime \prime }$ is defined as 
\begin{equation}
\sum_{\{i_n\}}{}^{\prime \prime }\equiv \underbrace{\sum_{i_1=1}^{N-n+1}%
\sum_{i_2=2}^{N-n+2}\cdots \sum_{i_{n-1}=n-1}^{N-1}\sum_{i_n=n}^N}%
_{\{i_1<i_2<\cdots <i_{n-1}<i_n\}}.  \label{8}
\end{equation}
The summation of Eq. (\ref{7}) includes $\binom Nn=N(N-1)\cdots (N-n+1)/n!$\
terms. Now, we define a normalized atomic storage state 
\begin{eqnarray}
|C_k^n\rangle &=&\sqrt{\frac{n!}{N(N-1)\cdots (N-n+1)}}  \label{9} \\
&&\times \sum_{\{i_n\}}{}^{\prime \prime }|b_1\cdots c_{i_1}\cdots
c_{i_n}\cdots b_N\rangle \exp [ik(z_{i_1}+\cdots +z_{i_n})].  \nonumber
\end{eqnarray}
Obviously, the atomic storage states with a different number of excitations
are orthogonal to each other 
\begin{equation}
\langle C_k^n|C_k^m\rangle =\delta _{nm}.  \label{9p}
\end{equation}
In this definition, the superposition state of $N$ collective atoms includes
any possible combination of $n$\ atoms being in the level $|c\rangle $,
while the corresponding spatial coherence is recorded in the phase of the
wavefunction. Physically, it means that $n$\ photons can be stored by any
combination of excited $n$\ atoms with an equal possibility, in
correspondence with the nonlocality for photons. However, the coherent
information of the field has been retained in the probability amplitudes.
Note that the atomic storage state, with a definite wavevector $k,$ is
independent of position $z,$\ which disappears in the summation.

By using definition (\ref{9}), Eq. (\ref{7}) becomes 
\begin{equation}
(\sigma _k^{\dagger })^n|C^0\rangle =\sqrt{\frac{N(N-1)\cdots (N-n+1)}{N^n}}%
\sqrt{n!}|C_k^n\rangle .  \label{10}
\end{equation}
It is easy to check that 
\begin{equation}
\sigma _k^{\dagger }|C_k^n\rangle =\sqrt{1-\frac nN}\sqrt{n+1}%
|C_k^{n+1}\rangle .  \label{11}
\end{equation}
The above two equations are exact. However, in the limit $N>>n$, the
corresponding approximate expressions are 
\begin{equation}
(\sigma _k^{\dagger })^n|C^0\rangle \simeq \sqrt{n!}|C_k^n\rangle ,
\label{12}
\end{equation}
and 
\begin{equation}
\sigma _k^{\dagger }|C_k^n\rangle \simeq \sqrt{n+1}|C_k^{n+1}\rangle .
\label{13}
\end{equation}

The annihilation operator is applied to the ''vacuum''state

\begin{equation}
\sigma _k|C^0\rangle =0.  \label{14}
\end{equation}
In Appendix A, the general formula for the annihilation operator is proved as

\begin{equation}
\sigma _k|C_k^n\rangle =\sqrt{1-\frac{n-1}N}\sqrt{n}|C_k^{n-1}\rangle \simeq 
\sqrt{n}|C_k^{n-1}\rangle .  \label{15}
\end{equation}
Equations (\ref{11}) and (\ref{15}) give immediately

\begin{mathletters}
\label{16}
\begin{eqnarray}
\sigma _k^{\dagger }\sigma _k|C_k^n\rangle  &=&(1-\frac{n-1}N)n|C_k^n\rangle
\simeq n|C_k^n\rangle ,  \label{16a} \\
\sigma _k\sigma _k^{\dagger }|C_k^n\rangle  &=&(1-\frac
nN)(n+1)|C_k^n\rangle \simeq (n+1)|C_k^n\rangle .  \label{16b}
\end{eqnarray}
The approximations in Eqs. (\ref{15}) and (\ref{16}) are valid in the limit $%
N>>n$.\ Equation (\ref{16}) verifies again the bosonic commutation in this
limit. If one admits both the bosonic commutation (\ref{4}) and Eq. (\ref{13}%
), by using the commutation 
\end{mathletters}
\begin{equation}
\lbrack \sigma _k,(\sigma _k^{\dagger })^n]\simeq n(\sigma _k^{\dagger
})^{n-1},  \label{17p}
\end{equation}
it can also obtain 
\begin{equation}
\sigma _k|C_k^n\rangle \simeq \sqrt{n}|C_k^{n-1}\rangle .  \label{17}
\end{equation}
The atomic storage states are also the eigenstates of the population
operators 
\begin{mathletters}
\label{17q}
\begin{eqnarray}
\sum_{j=1}^N(|b_j\rangle \langle b_j|C_k^n\rangle  &=&(N-n)|C_k^n\rangle ,
\label{17qa} \\
\sum_{j=1}^N(|c_j\rangle \langle c_j|C_k^n\rangle  &=&n|C_k^n\rangle .
\label{17qb}
\end{eqnarray}
(see Appendix A)

According to Dicke's definition ( Eq. (47) in Ref. \cite{dicke} ), the total
angular momentum operators of the atomic ensemble can be described as 
\end{mathletters}
\begin{mathletters}
\label{d1}
\begin{eqnarray}
R_{k1} &=&(\sqrt{N}/2)(\sigma _k^{\dagger }+\sigma _k),  \label{d1a} \\
R_{k2} &=&(-i\sqrt{N}/2)(\sigma _k^{\dagger }-\sigma _k),  \label{d1b} \\
R_3 &=&(N/2)(\sigma _k^{\dagger }\sigma _k-\sigma _k\sigma _k^{\dagger }),
\label{d1c} \\
R^2 &=&R_{k1}^2+R_{k2}^2+R_3^2  \label{d1d} \\
&=&(N/2)(\sigma _k^{\dagger }\sigma _k+\sigma _k\sigma _k^{\dagger
})+(N^2/4)(\sigma _k^{\dagger }\sigma _k-\sigma _k\sigma _k^{\dagger })^2. 
\nonumber
\end{eqnarray}
Using the exact relations of Eqs. (\ref{11}) and (\ref{15}), one obtains 
\end{mathletters}
\begin{mathletters}
\label{d2}
\begin{eqnarray}
R_3|C_k^n\rangle  &=&\frac 12(2n-N)|C_k^n\rangle ,  \label{d2a} \\
R^2|C_k^n\rangle  &=&\frac 12N(\frac 12N+1)|C_k^n\rangle .  \label{d2b}
\end{eqnarray}
Therefore, the atomic storage state defined in Eq. (\ref{9}) is the right
Dicke state with the maximum cooperation number $r=N/2$. The discussion in
this section exploits a new feature of Dicke state. The Dicke states with
the maximum cooperation number play the role of number states in front of
the collective lower and upper atomic operators.

\section{Multimode atomic storage states}

The multimode case is concerned with how the information of the multimode
photons is distributed in the local atomic excitations $|c_j\rangle .$ To
see it, we firstly consider a simple case --- the multimode
single-excitation atomic storage state; that is, each mode contains only one
excitation. We apply the multimode creation operators to the ''vacuum''
state 
\end{mathletters}
\begin{eqnarray}
&&\sigma _{k_1}^{\dagger }\cdots \sigma _{k_n}^{\dagger }|C^0\rangle 
\label{18} \\
&=&\frac 1{\sqrt{N^n}}\sum_{i_1=1}^N|c_{i_1}\rangle \langle b_{i_1}|\exp
(ik_1z_{i_1})\cdots \sum_{i_n=1}^N|c_{i_n}\rangle \langle b_{i_n}|\exp
(ik_nz_{i_n})|b_1b_2\cdots b_N\rangle   \nonumber \\
&=&\frac 1{\sqrt{N^n}}\sum_{\{i_n\}}{}^{\prime }|b_1\cdots c_{i_1}\cdots
c_{i_n}\cdots b_N\rangle \exp [i(k_1z_{i_1}+\cdots +k_nz_{i_n})].  \nonumber
\end{eqnarray}
This equation is apparently different to Eq. (\ref{6}) by the phase factors.
Indeed, the exchanges of the indices in the summation contribute to the same
atomic state, but, with different phase distributions. For example, $%
(i_1=1,i_2=2,i_3,\cdots ,i_n)$ and $(i_1=2,i_2=1,i_3,\cdots ,i_n)$ display
the same state $|c_1c_2b_3\cdots c_{i_3}\cdots c_{i_n}\cdots b_N\rangle $ ,
but with the phase factors $\exp [i(k_1z_1+k_2z_2+k_3z_{i_3}+\cdots
+k_nz_{i_n})]$ and $\exp [i(k_1z_2+k_2z_1+k_3z_{i_3}+\cdots +k_nz_{i_n})]$,
respectively. Mathematically, for a given atomic collective state $%
|b_1\cdots c_{i_1}\cdots c_{i_n}\cdots b_N\rangle $, it allocates $n!$\
phase factors due to $n!$ permutations for $n$\ elements. This means that an
atom in the level $|c_j\rangle $, located at position $z_j$, records the
information of all the modes. Because the field is global, each atom in the
medium experiences the field coherence of all the modes, and, vice versa,
the field of each mode affects all the excited atoms.

To simplify the sign, we define 
\begin{equation}
\{k_n\}\cdot \{z_{i_n}\}_l\equiv (k_1z_{i_1}+\cdots +k_nz_{i_n})_l,
\label{19}
\end{equation}
where $\{z_{i_n}\}_l$ stands for the $l$-th sequence of all the $n!$\
permutations for $n$ elements. Accordingly, Eq. (\ref{18}) can be written as 
\begin{equation}
\sigma _{k_1}^{\dagger }\cdots \sigma _{k_n}^{\dagger }|C^0\rangle =\frac 1{%
\sqrt{N^n}}\sum_{\{i_n\}}{}^{\prime \prime }|b_1\cdots c_{i_1}\cdots
c_{i_n}\cdots b_N\rangle \sum_{l=1}^{n!}\exp [i\{k_n\}\cdot \{z_{i_n}\}_l],
\label{20}
\end{equation}
where $\sum_{\{i_n\}}^{\prime \prime }$ has already been defined in Eq. (\ref
{8}). In comparison with the single mode case, shown in Eq. (\ref{7}), $n$%
-excitations in Eq. (\ref{20}) share the phases of $n$ modes. We define a
multimode single-excitation atomic storage state as 
\begin{eqnarray}
&&|C_{k_1}^1\cdots C_{k_n}^1\rangle \equiv \frac 1{\sqrt{N(N-1)\cdots (N-n+1)%
}}  \label{21} \\
&&\times \sum_{\{i_n\}}{}^{\prime \prime }|b_1\cdots c_{i_1}\cdots
c_{i_n}\cdots b_N\rangle \sum_{l=1}^{n!}\exp [i\{k_n\}\cdot \{z_{i_n}\}_l], 
\nonumber
\end{eqnarray}
which has been normalized, as shown in Appendix B. With the combination of
Eqs. (\ref{20}) and (\ref{21}), we obtain 
\begin{equation}
\sigma _{k_1}^{\dagger }\cdots \sigma _{k_n}^{\dagger }|C^0\rangle =\sqrt{%
\frac{N(N-1)\cdots (N-n+1)}{N^n}}|C_{k_1}^1\cdots C_{k_n}^1\rangle \simeq
|C_{k_1}^1\cdots C_{k_n}^1\rangle .  \label{22}
\end{equation}

Now, we discuss the general case of multimode atomic storage states. $s$\
modes containing total $n$ excitations can be generated by 
\begin{eqnarray}
&&(\sigma _{k_1}^{\dagger })^{m_1}\cdots (\sigma _{k_s}^{\dagger
})^{m_s}|C^0\rangle  \label{23} \\
&=&\frac 1{\sqrt{N^n}}\left( \sum_{j=1}^N|c_j\rangle \langle b_j|\exp
(ik_1z_j)\right) ^{m_1}\cdots \left( \sum_{j=1}^N|c_j\rangle \langle
b_j|\exp (ik_sz_j)\right) ^{m_s}|b_1b_2\cdots b_N\rangle  \nonumber
\end{eqnarray}
where $m_1+\cdots +m_s=n$. By defining an index set as 
\begin{equation}
\{i_n\}\equiv (i_1,\cdots ,i_{m_1},i_{m_1+1},\cdots ,i_{m_2},\cdots
,i_{n-m_s+1},\cdots ,i_n),  \label{24}
\end{equation}
Eq. (\ref{23}) can be written as 
\begin{eqnarray}
&&(\sigma _{k_1}^{\dagger })^{m_1}\cdots (\sigma _{k_s}^{\dagger
})^{m_s}|C^0\rangle  \label{25} \\
&=&\frac 1{\sqrt{N^n}}\sum_{\{i_n\}}{}^{\prime }|b_1\cdots c_{i_1}\cdots
c_{i_n}\cdots b_N\rangle  \nonumber \\
&&\times \exp [ik_1(z_{i_1}+\cdots +z_{i_{m_1}})+\cdots
+ik_s(z_{i_{n-m_s+1}}+\cdots +z_{i_n})].  \nonumber
\end{eqnarray}
Similarly, as in the previous cases, any particular atomic collective state $%
|b_1\cdots c_{j_1}\cdots c_{j_n}\cdots b_N\rangle $\ is related to $n!$\
terms in the summation throughout all indices. But, among these $n!$ terms,
the phase factor of each term will repeatedly appear $m_1!\cdots m_s!$ times
because exchanges of indices within a mode cause no difference. As a result,
the remaining non-repeated phase factors terms are $n!/(m_1!\cdots m_s!)$.
We define again 
\begin{equation}
\{k_s^{(m_s)}\}\cdot \{z_{i_n}\}_l\equiv \left( k_1(z_{i_1}+\cdots
+z_{i_{m_1}})+\cdots +k_s(z_{i_{n-m_s+1}}+\cdots +z_{i_n})\right) _l
\label{26}
\end{equation}
as one of these\ combinations with index $l$. Equation (\ref{25}) is written
as 
\begin{eqnarray}
&&(\sigma _{k_1}^{\dagger })^{m_1}\cdots (\sigma _{k_s}^{\dagger
})^{m_s}|C^0\rangle  \label{27} \\
&=&\frac{m_1!\cdots m_s!}{\sqrt{N^n}}\sum_{\{i_n\}}{}^{\prime \prime
}|b_1\cdots c_{i_1}\cdots c_{i_n}\cdots b_N\rangle
\sum_{l=1}^{n!/(m_1!\cdots m_s!)}\exp [i\{k_s^{(m_s)}\}\cdot \{z_{i_n}\}_l].
\nonumber
\end{eqnarray}
A general multimode atomic storage state can be defined as 
\begin{eqnarray}
&&|C_{k_1}^{m_1}\cdots C_{k_s}^{m_s}\rangle  \label{28} \\
&\equiv &\sqrt{\frac{m_1!\cdots m_s!}{N(N-1)\cdots (N-n+1)}}  \nonumber \\
&&\times \sum_{\{i_n\}}{}^{\prime \prime }|b_1\cdots c_{i_1}\cdots
c_{i_n}\cdots b_N\rangle \sum_{l=1}^{n!/(m_1!\cdots m_s!)}\exp
[i\{k_s^{(m_s)}\}\cdot \{z_{i_n}\}_l],  \nonumber
\end{eqnarray}
which has been normalized (see Appendix B). Finally, Eq. (\ref{27}) becomes 
\begin{eqnarray}
&&(\sigma _{k_1}^{\dagger })^{m_1}\cdots (\sigma _{k_s}^{\dagger
})^{m_s}|C^0\rangle  \label{29} \\
&=&\sqrt{\frac{N(N-1)\cdots (N-n+1)}{N^n}}\sqrt{m_1!\cdots m_s!}%
|C_{k_1}^{m_1}\cdots C_{k_s}^{m_s}\rangle  \nonumber \\
&\simeq &\sqrt{m_1!\cdots m_s!}|C_{k_1}^{m_1}\cdots C_{k_s}^{m_s}\rangle , 
\nonumber
\end{eqnarray}
where the approximation is valid in the limit$\ N>>n$. In this limit, the
creation of an excitation of mode $l$\ for a multimode storage state is
written as 
\begin{equation}
\sigma _{k_l}^{\dagger }|C_{k_1}^{m_1}\cdots C_{k_l}^{m_l}\cdots
C_{k_s}^{m_s}\rangle \simeq \sqrt{m_l+1}|C_{k_1}^{m_1}\cdots
C_{k_l}^{m_l+1}\cdots C_{k_s}^{m_s}\rangle .  \label{31}
\end{equation}
Similarly, as in the single mode case, by considering the bosonic
commutation (\ref{4}) and Eq. (\ref{29}), the annihilation of an excitation
of mode $l$ is written as 
\begin{equation}
\sigma _{k_l}|C_{k_1}^{m_1}\cdots C_{k_l}^{m_l}\cdots C_{k_s}^{m_s}\rangle
\simeq \sqrt{m_l}|C_{k_1}^{m_1}\cdots C_{k_l}^{m_l-1}\cdots
C_{k_s}^{m_s}\rangle .  \label{32}
\end{equation}

Equations (\ref{10}), (\ref{22}) and (\ref{29}) imply that the state $%
|C^0\rangle $\ defined in Eq. (\ref{5}) represents a vacuum state not only
for a single mode but also for a multimode. Physically, it shows that the
ultracold collective atoms are able to store a multimode field. As a result,
we may derive 
\begin{equation}
\sigma _{k_2}^{\dagger }|C_{k_1}^{m_1}\rangle \simeq
|C_{k_2}C_{k_1}^{m_1}\rangle ,  \label{33}
\end{equation}
where $|C_{k_1}^{m_1}\rangle $ can be understood as either a single mode
state or a multimode state with the vacuum for those modes other than mode $%
k_1$.

In the above theoretical description, we have shown that the atomic storage
states are a duplicate of the Fock states of the electromagnetic field. The
explicit expressions of atomic storage states Eqs. (\ref{9}) and (\ref{28})
display duality of particle and coherence. The excitations may appear
everywhere with an equal probability in the medium, corresponding with the
nonlocality for photons. However, each excitation takes the local phase
factors of the single-mode, or multimode, fields as the quantum probability
amplitude, therefore recording the coherence. Consequently, it may establish
the correspondence of two quantum system, the field and the atomic ensemble.
This provides the basis for complete storage of quantum information of the
bosonic field in an atomic medium.

\section{Dark states in EIT}

In the EIT configuration, the weak signal field interacting resonantly with
the atomic transition $|a\rangle -|b\rangle $ is described by the field
operator 
\begin{eqnarray}
E_s(z,t) &=&(1/2)\mathcal{E}_0a(z)\exp [i(k_sz-\omega _st)]+\text{h.c.}
\label{34} \\
&=&(1/2)\mathcal{E}_0\sum_qa(q)\exp [iqz]\exp [i(k_sz-\omega _st)]+\text{h.c.%
},  \nonumber
\end{eqnarray}
where $\mathcal{E}_0$ is the field amplitude per photon and $\omega _s=ck_s$%
. $c$\ is the speed of light in a vacuum. The strong control field,
resonantly driving the atomic transition $|a\rangle -|c\rangle $ is assumed
as classical 
\begin{equation}
E_c(z,t)=\frac{\hbar \Omega }{2\wp _{ac}}\exp [i(k_cz-\omega _ct)]+\text{%
c.c.,}  \label{35}
\end{equation}
where $\wp _{ac}$ is the dipole moment of the transition $|a\rangle
-|c\rangle $ and $\omega _c=ck_c$. In the interaction picture, the
interaction Hamiltonian is described as 
\begin{eqnarray}
H_I &=&\hbar \sum_q\omega _qa^{\dagger }(q)a(q)  \label{36} \\
&&-\frac \hbar 2\sum_{j=1}^N\left\{ g\sum_qa(q)|a_j\rangle \langle b_j|\exp
[i(k_s+q)z_j]+\Omega |a_j\rangle \langle c_j|\exp [ik_cz_j]+\text{h.c.}%
\right\} ,  \nonumber
\end{eqnarray}
where $\omega _q=cq$ is the detuning of mode $q$ with respect to the
resonant frequency $\omega _s$ of the signal field. By defining the atomic
collective operators 
\begin{mathletters}
\label{37}
\begin{eqnarray}
\rho _{ab}(q) &=&\frac 1N\sum_{j=1}^N|a_j\rangle \langle b_j|\exp
[i(k_s+q)z_j],  \label{37a} \\
\rho _{ac}(q) &=&\frac 1N\sum_{j=1}^N|a_j\rangle \langle c_j|\exp
[i(k_c+q)z_j],  \label{37b}
\end{eqnarray}
the Hamiltonian (\ref{36}) is written as 
\end{mathletters}
\begin{equation}
H_I=\hbar \sum_q\omega _qa^{\dagger }(q)a(q)-\frac \hbar 2\left\{
gN\sum_qa(q)\rho _{ab}(q)+\Omega N\rho _{ac}(0)+\text{h.c.}\right\} .
\label{38}
\end{equation}

The new quantum field operator defined in Ref. \cite{flei} is written as 
\begin{equation}
\psi _q=\cos \theta a_q-\sin \theta \sigma _q,  \label{39}
\end{equation}
where 
\begin{equation}
\cos \theta =\Omega /\sqrt{\Omega ^2+g^2N},\qquad \sin \theta =g\sqrt{N}/%
\sqrt{\Omega ^2+g^2N}.  \label{40}
\end{equation}
The transition $|b\rangle -|c\rangle $\ concerns both the absorption of a
signal photon and the emission of a driving photon. Replacing $k$ by $%
(k_s+q)-k_c$\ in Eq. (\ref{2}), one obtains the annihilation operator $%
\sigma _q$\ of the collective atoms in EIT 
\begin{equation}
\sigma _q=\frac 1{\sqrt{N}}\sum_{j=1}^N|b_j\rangle \langle c_j|\exp
[-i(k_s+q-k_c)z_j].  \label{41}
\end{equation}
Correspondingly, $k$ should also be replaced by $k_s+q-k_c$ in the atomic
storage states. $\psi _q$ satisfies the bosonian commutation relation as
long as $\sigma _q$\ does 
\begin{equation}
\lbrack \psi _q,\psi _{q^{\prime }}^{\dagger }]=\cos ^2\theta
[a_q,a_{q^{\prime }}^{\dagger }]+\sin ^2\theta [\sigma _q,\sigma _{q^{\prime
}}^{\dagger }]\simeq \delta _{qq^{\prime }}.  \label{42}
\end{equation}
It has been shown in Eq. (\ref{40}), that the parameter $\theta $ is related
to the strength of the control field. In the strong and weak limits of the
control field, $\psi _q$ tends to $a_q$ and $\sigma _q$, respectively.

According to Ref. \cite{flei}, the dark state is defined as 
\begin{equation}
|D_q^n\rangle =\frac 1{\sqrt{n!}}(\psi _q^{\dagger })^n|0\rangle |C^0\rangle
,  \label{43}
\end{equation}
where $|0\rangle $ is the vacuum state of the signal field. The lowest dark
state is designated as $|D^0\rangle \equiv |0\rangle |C^0\rangle $. The
quasi-particle in the dark state is called a polariton \cite{flei}.

Using Eq. (\ref{10}), one obtains the exact expression of the dark state 
\begin{eqnarray}
|D_q^n\rangle &=&\sum_{m=0}^n(-1)^m\sqrt{\frac{n(n-1)\cdots (n-m+1)}{m!}}%
\sqrt{\frac{N(N-1)\cdots (N-m+1)}{N^m}}  \nonumber  \label{44} \\
&&\times \cos ^{n-m}\theta \sin ^m\theta |n-m\rangle |C_q^m\rangle
\label{44} \\
&=&\sum_{m=0}^n\sqrt{\frac{n(n-1)\cdots (n-m+1)}{m!}}\sqrt{\frac{%
N(N-1)\cdots (N-m+1)}{N^m}}  \nonumber \\
&&\times \frac{\Omega ^{n-m}(-g\sqrt{N})^m}{(\Omega ^2+g^2N)^{n/2}}%
|n-m\rangle |C_q^m\rangle .  \nonumber
\end{eqnarray}
The dark states described above are orthogonal to each other since they have
different quasiparticle numbers, but are not normalized. Under the condition 
$N>>n$, the dark state can be approximately written as 
\begin{eqnarray}
|D_q^n\rangle &\simeq &\sum_{m=0}^n(-1)^m\sqrt{\frac{n(n-1)\cdots (n-m+1)}{m!%
}}\cos ^{n-m}\theta \sin ^m\theta |n-m\rangle |C_q^m\rangle  \nonumber \\
&=&\sum_{m=0}^n\sqrt{\frac{n(n-1)\cdots (n-m+1)}{m!}}\frac{\Omega ^{n-m}(-g%
\sqrt{N})^m}{(\Omega ^2+g^2N)^{n/2}}|n-m\rangle |C_q^m\rangle .  \label{45}
\end{eqnarray}
The above expression of the dark state satisfies the normalized orthogonal
relation 
\begin{equation}
\langle D_q^n|D_q^m\rangle =\delta _{nm}.  \label{46}
\end{equation}
Eq. (\ref{45}) shows that when the parameter $\theta $ is taken to be $0$\
and $\pi /2$, the summation in the dark states reduces to only the first and
the last term 
\begin{mathletters}
\label{46p}
\begin{eqnarray}
|D_q^n\rangle &=&|n\rangle |C^0\rangle \qquad \text{for \ \ }\theta =0,
\label{46a} \\
|D_q^n\rangle &=&(-1)^n|0\rangle |C_q^n\rangle \qquad \text{for \ \ }\theta
=\pi /2,  \label{46b}
\end{eqnarray}
respectively. Therefore, by varying $\theta $ adiabatically, the
quasi-particles can be transferred between the photon state and the atomic
storage state.

According to definition (\ref{43}), one can obtain the exact expression 
\end{mathletters}
\begin{equation}
\psi _q^{\dagger }|D_q^n\rangle =\frac 1{\sqrt{n!}}(\psi _q^{\dagger
})^{n+1}|0\rangle |C^0\rangle =\sqrt{n+1}|D_q^{n+1}\rangle .  \label{47}
\end{equation}
It is easy to check 
\begin{equation}
\psi _q|D^0\rangle =0.  \label{48}
\end{equation}
As the same for the operator $\sigma _q$, with the help of the bosonic
commutation relation (\ref{42}), one obtains for the dark state (\ref{45}) 
\begin{equation}
\psi _q|D_q^n\rangle \simeq \sqrt{n}|D_q^{n-1}\rangle .  \label{49}
\end{equation}
Moreover, the multimode dark state can be generated by 
\begin{equation}
|D_{q_1}^{n_1}\cdots D_{q_s}^{n_s}\rangle =\frac 1{\sqrt{n_1!\cdots n_s!}%
}(\psi _{q_1}^{\dagger })^{n_1}\cdots (\psi _{q_s}^{\dagger
})^{n_s}|0\rangle |C^0\rangle .  \label{51}
\end{equation}
They can be treated just like the multimode photon number states.

In Appendix C, we have proven that, at the exact resonance, both the exact
expression (\ref{44}) and the approximate expression (\ref{45}) of the dark
states are the eigenstates of the interaction Hamiltonian (\ref{38}) with a
null eigenvalue. A pulse of monochromatic light has a narrow bandwidth, and
the detuning $\omega _q$\ from the carrier frequency $\omega _s$\ is small.
If we omit the first term in the Hamiltonian (\ref{38}), the multimode dark
states consisting of the pulse are the eigenstates of the interaction
Hamiltonian.

Assume that, at the initial time, a signal pulse is at a multimode state 
\begin{equation}
\sum_{\{q_s\}}\alpha (q_1,\cdots ,q_s)|n_1\cdots n_s\rangle ,  \label{52}
\end{equation}
while the cold collective atoms are, approximately, in the ground state $%
|C^0\rangle $. The combined system of the signal field and the atoms is in
the state 
\begin{equation}
|\Psi (0)\rangle =\sum_{\{q_s\}}\alpha (q_1,\cdots ,q_s)|n_1\cdots
n_s\rangle |C^0\rangle =\sum_{\{q_s\}}\alpha (q_1,\cdots
,q_s)|D_{q_1}^{n_1}\cdots D_{q_s}^{n_s}\rangle _{\theta =0}.  \label{53}
\end{equation}
When the control field is strong enough, the signal pulse can maintain and
transmit through the medium. Note that $|\Psi (0)\rangle $ is also the
eigenstate of the interaction Hamiltonian with a null eigenvalue. If the
control field is changed adiabatically to a very small level at a later time 
$t_1$, the state of the system is also changed adiabatically to 
\begin{eqnarray}
|\Psi (t_1)\rangle &=&\sum_{\{q_s\}}\alpha (q_1,\cdots
,q_s)|D_{q_1}^{n_1}\cdots D_{q_s}^{n_s}\rangle _{\theta =\pi /2}  \label{54}
\\
&=&\sum_{\{q_s\}}(-1)^{n_1+\cdots +n_s}\alpha (q_1,\cdots ,q_s)|0\rangle
|C_{q_1}^{n_1}\cdots C_{q_s}^{n_s}\rangle .  \nonumber
\end{eqnarray}
It forms an associate state for $|\Psi (0)\rangle $. The whole of the
quantum information of the signal pulse has been stored in the atomic
medium, in the form of a ''negative copy'', in which each excitation changes
a $\pi $-phase. As soon as the control field returns to the previous level,
the state (\ref{53}) is recovered. Conversely, if Eq. (\ref{54}) is an
initial state generated in other model, by turning on the control field, it
will be converted to the corresponding optical field, enabling it to be seen.

\section{Dynamic quantum transfer in macroscopic matter}

Due to the fact that the dark states are eigenstates of the EIT interaction,
quantum transfer processes between field and matter are quasi-stationary by
adiabatically changing the control field. On the other hand, the transfer
can be performed in a dynamic way, which has been described in the
literature \cite{pol}, \cite{mat}. In this section, we study a general
description for dynamic transfer of quantum state between field and
ultracold matter. The interaction configuration can be designed as, either,
the parametric process or the Raman transition \cite{pol}, in which both a
weak signal beam and a strong control beam interact resonantly with two
levels of atoms.

For simplicity, we consider a single-mode interaction. In the interaction
picture, the effective interaction Hamiltonian is written as 
\begin{equation}
H_I=\hbar \Omega (a\sigma ^{\dagger }+a^{\dagger }\sigma ),  \label{55}
\end{equation}
where $\Omega $ is the Rabi frequency of the control beam, assumed as
classical. The collective atomic operator $\sigma $ is defined by Eq. (\ref
{2p}) and behaves boson-like in the low excitation limit. The model is well
known as a coupled harmonic oscillator, and can be solved exactly. Here we
illustrate this model again from a new viewpoint by introducing a very
simple method for the exact solution of the state-vector evolution. For this
model, it is easy to obtain the evolution of the operators in the Heisenberg
picture, 
\begin{equation}
\binom{a(t)}{\sigma (t)}=\left( 
\begin{array}{cc}
\cos \Omega t & -i\sin \Omega t \\ 
-i\sin \Omega t & \cos \Omega t
\end{array}
\right) \binom{a(0)}{\sigma (0)}.  \label{56}
\end{equation}
With this method, if the initial state can be written as $|\Psi (0)\rangle
=f(a(0),\sigma (0))|\Theta (0)\rangle ,$ while the evolution of the state $%
|\Theta (0)\rangle $ is already known to be $|\Theta (t)\rangle ,$ we obtain 
\begin{eqnarray}
|\Psi (t)\rangle &=&U(t)|\Psi (0)\rangle =U(t)f(a(0),\sigma (0))|\Theta
(0)\rangle  \label{57} \\
&=&U(t)f(a(0),\sigma (0))U^{-1}(t)U(t)|\Theta (0)\rangle =f(a(-t),\sigma
(-t))|\Theta (t)\rangle ,  \nonumber
\end{eqnarray}
where $U(t)=\exp (-iH_It/\hbar )$. An initial Fock state for the signal
photons $m$ and the atomic collective excitations $n$ is represented by $%
|m,C^n\rangle =(1/\sqrt{m!n!})[a^{\dagger }(0)]^m[\sigma ^{\dagger
}(0)]^n|0,C^0\rangle $. Since the evolution of the vacuum state, as $|\Theta
(0)\rangle $ in Eq. (\ref{57}), is known to be $|\Theta (t)\rangle =\exp
(-iH_It/\hbar )|0,C^0\rangle =|0,C^0\rangle ,$ we obtain the evolution for
an initial Fock state $|m,C^n\rangle $ 
\begin{equation}
|\Psi _{mn}(t)\rangle =(1/\sqrt{m!n!})[a^{\dagger }(-t)]^m[\sigma ^{\dagger
}(-t)]^n|0,C^0\rangle ,  \label{58}
\end{equation}
where the dynamical operators are written as 
\begin{mathletters}
\label{58p}
\begin{eqnarray}
a^{\dagger }(-t) &=&U(t)a^{\dagger }U^{-1}(t)=a^{\dagger }\cos \Omega
t-i\sigma ^{\dagger }\sin \Omega t,  \label{58pa} \\
\sigma ^{\dagger }(-t) &=&U(t)\sigma ^{\dagger }U^{-1}(t)=\sigma ^{\dagger
}\cos \Omega t-ia^{\dagger }\sin \Omega t,  \label{58pb}
\end{eqnarray}
where $a^{\dagger }\equiv a^{\dagger }(0)$ and $\sigma ^{\dagger }\equiv
\sigma ^{\dagger }(0)$. According to Eq. (\ref{58}), the exact evolution for
an arbitrary initial state $\sum \xi _{mn}|m,C^n\rangle $ is therefore $\sum
\xi _{mn}|\Psi _{mn}(t)\rangle .$

The state $|\Psi _{mn}(t)\rangle $ described by Eq. (\ref{58}) evolves
periodically with the fundamental frequency $\Omega $ and conserves the
total particle number. For example, in the simplest case of only a single
excitation $|1,C^0\rangle $, the time evolution is $|\Psi _{10}(t)\rangle
=\cos \Omega t|1,C^0\rangle -i\sin \Omega t|0,C^1\rangle $. The entanglement
between two subsystems increases in time and reaches a maximum at $\Omega
t=\pi /4$. Then, the entanglement decreases and the excitation transfers
completely from one subsystem to another at $\Omega t=\pi /2$. In general,
if the initial state has $m$ signal photons and no excitation for atoms $%
|m,C^0\rangle $, the time evolution is written as 
\end{mathletters}
\begin{eqnarray}
|\Psi _{m0}(t)\rangle  &=&(1/\sqrt{m!})[a^{\dagger }(-t)]^m|0,C^0\rangle =(1/%
\sqrt{m!})[a^{\dagger }\cos \Omega t-i\sigma ^{\dagger }\sin \Omega
t]^m|0,C^0\rangle   \nonumber  \label{59} \\
&=&\sum_{j=0}^m(-i)^j\sqrt{\frac{m!}{j!(m-j)!}}(\cos \Omega t)^{m-j}(\sin
\Omega t)^j|m-j,C^j\rangle .  \label{59}
\end{eqnarray}
The evolution states can be called ''dynamic polaritons'', since they are
compatible with the ''stationary polaritons'' defined by Eq. (\ref{45}). At
the times $\Omega t=(1/2)\pi ,$ $\pi $ and $(3/2)\pi $, the evolution state
has been de-entangled to $(-i)^m|0,C^m\rangle $, $(-1)^m|m,C^0\rangle ,$ and 
$i^m|0,C^m\rangle $, respectively. This means, at certain times, the
entanglement formed in the dynamical process can be cancelled and the
excitations are transferred completely from one subsystem to another. When
an initial state is an arbitrary superposition of Fock states for the signal
beam and the ''vacuum'' for the ultracold atoms, i. e. $|\Psi (0)\rangle
=|\Phi ,C^0\rangle $ where $|\Phi \rangle =\sum \alpha _m|m\rangle $, it
evolves to the states $|0,\Phi ^{(-i)}\rangle ,|\Phi ^{(-)},C^0\rangle ,$
and $|0,\Phi ^{(i)}\rangle $ at the times $\Omega t=(1/2)\pi ,$ $\pi $ and $%
(3/2)\pi $, respectively, and comes back to the original at $\Omega t=2\pi .$
Here we define three states associated with $|\Phi \rangle =\sum \alpha
_m|m\rangle $ 
\begin{equation}
|\Phi ^{(\pm i)}\rangle =\sum (\pm i)^m\alpha _m|m\rangle ,\qquad |\Phi
^{(-)}\rangle =\sum (-1)^m\alpha _m|m\rangle .  \label{60}
\end{equation}
Obviously, these associate states have the same particle distribution, but
with different phase shifts in amplitudes. As a matter of fact, the phase
factor in the superposition can be observed macroscopically. For instance,
if $|\Phi \rangle $ is a coherent state $|\alpha \rangle $, one obtains the
coherent states again for the associate states with a particular phase
shift, i. e. $|\Phi ^{(\pm i)}\rangle =|\pm i\alpha \rangle $ and $|\Phi
^{(-)}\rangle =|-\alpha \rangle .$ As for an arbitrary state, it is also
true by means of the expectation value of the amplitude operator, such that $%
\langle \Phi ^{(\pm i)}|a|\Phi ^{(\pm i)}\rangle =\pm i\langle \Phi |a|\Phi
\rangle $ and $\langle \Phi ^{(-)}|a|\Phi ^{(-)}\rangle =-\langle \Phi
|a|\Phi \rangle $. Though, in general, these associate states are not
identical to the original, owing to a phase shift; the quantum information
of the original state can still be faithfully preserved. It looks like a
photograph and the corresponding negative copy. The associate states $|\Phi
^{(\pm i)}\rangle $ and $|\Phi ^{(-)}\rangle $ can be seen as ''orthogonal''
and ''negative'' copies of a quantum ''picture'' $|\Phi \rangle $.

In the dynamic quantum transfer, a quantum state of the signal field can be
stored in and then retrieved from a medium by turning off and on the control
field at a certain time. Because the dynamic polariton state defined by Eq. (%
\ref{59}) is the eigenstate of the free Hamiltonian, it will be preserved
while the interaction is turning off. Similarly, the model can be extended
to the multimode case provided the conditions for the radiation wavelength
Eq. (\ref{c}) is satisfied. Thus, coherent information of the optical field
can be transferred simultaneously.

Finally, we indicate that, for a proper transfer in this model, collective
atoms must be prepared in an atomic storage state. For ultracold matter, its
initial state can be considered as the vacuum state $|C^0\rangle $
approximately. If, at the initial time, the field is at an arbitrary state $%
|\Phi _1\rangle $ while collective atoms have been prepared in a
superposition state $|\Phi _2\rangle =\sum \beta _m|C^m\rangle $, $|\Phi
_1,\Phi _2\rangle $ will evolve to the corresponding states $|\Phi
_2^{(-i)},\Phi _1^{(-i)}\rangle ,|\Phi _1^{(-)},\Phi _2^{(-)}\rangle $ and $%
|\Phi _2^{(i)},\Phi _1^{(i)}\rangle $ at the certain times mentioned above.
It displays a complete swapping of quantum states for the coupled harmonic
oscillators.

\section{Conclusion}

In conclusion, we define collective atomic operators and atomic storage
states by containing spatial coherence and illustrate the conditions under
which the multimode collective atomic lower and upper operators are
boson-like. We indicate the fact that the atomic storage states shown by
definition (\ref{9}) are Dicke states with the maximum cooperation number.
The new feature for these Dicke states is that, in the low excitation limit
for a large number of atoms, they behave as the Fock states of an
electromagnetic field. The complete description and the deductive explicit
expressions for the atomic storage states present better physical
understanding of why the atomic ensemble can record fully the quantum
information, both the excitation and the coherence, of an optical
electromagnetic field. In addition to adiabatic quantum transfer by means of
dark states in EIT, we discuss the mechanism of dynamic quantum transfer via
dynamic polaritons which is formed by the fundamental interaction between
the field and ultracold matter. A combination of adiabatic and dynamic
schemes may find more applications in quantum information technology.

\section*{Acknowledgments}

This research was supported by the National Fundamental Research Project
(973) No. 2001CB309310 and the National Natural Science Foundation of China,
Project No. 10074008.

\section*{Appendix A}

First, we derive the exact equation (\ref{15}). By applying definitions (\ref
{2}) and (\ref{9}), it gives 
\begin{eqnarray}
\sigma _k|C_k^n\rangle  &=&\frac 1{\sqrt{N}}\sum_{l=1}^N|b_l\rangle \langle
c_l|\exp [-ikz_l]\sqrt{\frac{n!}{N\cdots (N-n+1)}}  \tag{A1}  \label{A1} \\
&&\times \sum_{\{i_n\}}{}^{\prime \prime }|b_1\cdots c_{i_1}\cdots
c_{i_n}\cdots b_N\rangle \exp [ik(z_{i_1}+\cdots +z_{i_n})].  \nonumber
\end{eqnarray}
When the operator $\sum_{l=1}^N|b_l\rangle \langle c_l|$ applies to a
particular state $|b_1\cdots c_{i_1}\cdots c_{i_n}\cdots b_N\rangle $, it
produces a superposition of $n$ states in which $|c_{i_k}\rangle $ is
sequentially replaced by $|b_{i_k}\rangle $. After this operation, Eq. (\ref
{A1}) is a summation of $n\times \binom Nn=N(N-1)\cdots (N-n+1)/(n-1)!$
states, in which $n-1$ atoms are populated in the level $|c\rangle $. It is
in fact $N-n+1$ times $|C_k^{n-1}\rangle $. For example, a particular state
in $|C_k^{n-1}\rangle $, say, $|c_1\cdots c_{n-1}b_n\cdots b_N\rangle ,$
comes from $N-(n-1)$ states, $|c_1\cdots c_{n-1}c_nb_{n+1}\cdots b_N\rangle ,
$ $|c_1\cdots c_{n-1}b_nc_{n+1}\cdots b_N\rangle ,$ $...,$ $|c_1\cdots
c_{n-1}b_nb_{n+1}\cdots c_N\rangle ,$ in $|C_k^n\rangle .$ Thus, Eq. (\ref
{A1}) is written as 
\begin{eqnarray}
\sigma _k|C_k^n\rangle  &=&\frac 1{\sqrt{N}}\sqrt{\frac{n!}{N\cdots (N-n+1)}}%
(N-n+1)  \tag{A2}  \label{A2} \\
&&\times \sum_{\{i_{n-1}\}}{}^{\prime \prime }|b_1\cdots c_{i_1}\cdots
c_{i_{n-1}}\cdots b_N\rangle \exp [ik(z_{i_1}+\cdots +z_{i_{n-1}})] 
\nonumber \\
&=&\frac 1{\sqrt{N}}\sqrt{\frac{n!}{N\cdots (N-n+1)}}(N-n+1)\sqrt{\frac{%
N\cdots (N-n+2)}{(n-1)!}}|C_k^{n-1}\rangle   \nonumber \\
&=&\sqrt{\frac{N-n+1}N}\sqrt{n}|C_k^{n-1}\rangle .  \nonumber
\end{eqnarray}

Then, we prove Eqs. (\ref{17q}). Equation (\ref{17qa}) is written as 
\begin{eqnarray}
&&\sum_{l=1}^N(|b_l\rangle \langle b_l|C_k^n\rangle   \tag{A3}  \label{A3} \\
&=&\sqrt{\frac{n!}{N\cdots (N-n+1)}}  \nonumber \\
&&\times \sum_{l=1}^N\sum_{\{i_n\}}{}^{\prime \prime }|b_l\rangle \langle
b_l|b_1\cdots c_{i_1}\cdots c_{i_n}\cdots b_N\rangle \exp [ik(z_{i_1}+\cdots
+z_{i_n})]  \nonumber \\
&=&\sqrt{\frac{n!}{N\cdots (N-n+1)}}\sum_{l=1}^N\sum_{\{i_n\}}{}^{\prime
\prime }|b_1\cdots c_{i_1}\cdots c_{i_n}\cdots b_N\rangle   \nonumber \\
&&\times \exp [ik(z_{i_1}+\cdots +z_{i_n})](1-\delta _{li_1})\cdots
(1-\delta _{li_n}).  \nonumber
\end{eqnarray}
Because all the indices of $i_k$ are not equal to one another, one has 
\begin{equation}
(1-\delta _{li_1})\cdots (1-\delta _{li_n})=1-(\delta _{li_1}+\cdots +\delta
_{li_n})+(\delta _{li_1}\delta _{li_2}+\cdots )-\cdots =1-(\delta
_{li_1}+\cdots +\delta _{li_n}).  \tag{A4}  \label{A4}
\end{equation}
Substituting Eq. (\ref{A4}) into Eq. (\ref{A3}), one obtains 
\begin{equation}
\sum_{l=1}^N(|b_l\rangle \langle b_l|C_k^n\rangle =(N-n)|C_k^n\rangle . 
\tag{A5}  \label{A5}
\end{equation}
Eq. (\ref{17qb}) is proved as 
\begin{eqnarray}
&&\sum_{l=1}^N(|c_l\rangle \langle c_l|C_k^n\rangle   \tag{A6}  \label{A6} \\
&=&\sqrt{\frac{n!}{N\cdots (N-n+1)}}  \nonumber \\
&&\times \sum_{l=1}^N\sum_{\{i_n\}}{}^{\prime \prime }|c_l\rangle \langle
c_l|b_1\cdots c_{i_1}\cdots c_{i_n}\cdots b_N\rangle \exp [ik(z_{i_1}+\cdots
+z_{i_n})]  \nonumber \\
&=&\sqrt{\frac{n!}{N\cdots (N-n+1)}}  \nonumber \\
&&\times \sum_{l=1}^N\sum_{\{i_n\}}{}^{\prime \prime }|b_1\cdots
c_{i_1}\cdots c_{i_n}\cdots b_N\rangle \exp [ik(z_{i_1}+\cdots
+z_{i_n})](\delta _{li_1}+\cdots +\delta _{li_n})  \nonumber \\
&=&n|C_k^n\rangle .  \nonumber  \label{A8}
\end{eqnarray}

\section*{Appendix B}

In this appendix, we calculate the normalized coefficient of the multimode
storage states. A multimode single-excitation state is defined as 
\begin{equation}
|C_{k_1}^1\cdots C_{k_n}^1\rangle \equiv \alpha \sum_{\{i_n\}}{}^{\prime
\prime }|b_1\cdots c_{i_1}\cdots c_{i_n}\cdots b_N\rangle
\sum_{l=1}^{n!}\exp [i\{k_n\}\cdot \{z_{i_n}\}_l].  \tag{B1}  \label{B1}
\end{equation}
The probability of each state in the above superposition is 
\begin{eqnarray}
\left| \sum_{l=1}^{n!}\exp [i\{k_n\}\cdot \{z_{i_n}\}_l]\right| ^2
&=&\sum_{l,j}^{n!}\exp [i\{k_n\}\cdot (\{z_{i_n}\}_l-\{z_{i_n}\}_j)] 
\tag{B2}  \label{B2} \\
&=&n!+\sum_{l\neq j}^{n!}\exp [i\{k_n\}\cdot (\{z_{i_n}\}_l-\{z_{i_n}\}_j)].
\nonumber
\end{eqnarray}
Then, we sum all these probabilities. For the first term of Eq. (\ref{B2}),
it is simply 
\begin{equation}
\sum_{\{i_n\}}{}^{\prime \prime }n!=n!\frac{N(N-1)\cdots (N-n+1)}{n!}%
=N(N-1)\cdots (N-n+1).  \tag{B3}  \label{B3}
\end{equation}
In the multimode case, one must find a mode with $k_j\neq 0$. By using Eq. (%
\ref{3t}), the summation $\sum {}_{\{i_n\}}^{\prime \prime }$ to the second
term of Eq. (\ref{B2}) vanishes. Therefore, we obtain 
\begin{equation}
\alpha =\frac 1{\sqrt{N(N-1)\cdots (N-n+1)}}.  \tag{B4}  \label{B4}
\end{equation}

Similarly, for a general multimode storage state defined by Eq. (\ref{28}),
the probability of finding a single state is 
\begin{eqnarray}
&&\left| \sum_{l=1}^{n!/(m_1!\cdots m_s!)}\exp [i\{k_s^{(m_s)}\}\cdot
\{z_{i_n}\}_l]\right| ^2  \tag{B5}  \label{B5} \\
&=&\sum_{l,j}^{n!/(m_1!\cdots m_s!)}\exp [i\{k_s^{(m_s)}\}\cdot
(\{z_{i_n}\}_l-\{z_{i_n}\}_j)]  \nonumber \\
&=&\frac{n!}{m_1!\cdots m_s!}+\sum_{l\neq j}^{n!/(m_1!\cdots m_s!)}\exp
[i\{k_s^{(m_s)}\}\cdot (\{z_{i_n}\}_l-\{z_{i_n}\}_j)].  \nonumber
\end{eqnarray}
The summation to the first term of the above equation gives 
\begin{eqnarray}
&&\sum_{\{i_n\}}{}^{\prime \prime }\frac{n!}{m_1!\cdots m_s!}  \tag{B6}
\label{B6} \\
&=&\frac{n!}{m_1!\cdots m_s!}\frac{N(N-1)\cdots (N-n+1)}{n!}=\frac{%
N(N-1)\cdots (N-n+1)}{m_1!\cdots m_s!}.  \nonumber
\end{eqnarray}
With the same reason, the summation to the second term vanishes. The
normalized coefficient is therefore 
\begin{equation}
\alpha =\sqrt{\frac{m_1!\cdots m_s!}{N(N-1)\cdots (N-n+1)}}.  \tag{B7}
\label{B7}
\end{equation}

\section*{Appendix C}

We define a new collective atomic state in which $n$ atoms are in the level $%
|c\rangle $ whereas one atom is in the level $|a\rangle $%
\begin{eqnarray}
|A_q^1,C_q^n\rangle &\equiv &\sqrt{\frac{n!}{N(N-1)\cdots (N-n)}}\sum_{l\neq
\{i_n\}}\sum_{\{i_n\}}{}^{\prime \prime }|b_1\cdots a_l\cdots c_{i_1}\cdots
c_{i_n}\cdots b_N\rangle  \nonumber  \label{C1} \\
&&\times \exp [i(k_s+q-k_c)(z_{i_1}+\cdots +z_{i_n})]\exp [i(k_s+q)z_l], 
\tag{C1}  \label{C1}
\end{eqnarray}
where $\sum_{l\neq \{i_n\}}$ designates the summation for index $l$ which
cannot be taken as $i_1,\cdots i_n$. We have already indicated that, the
state $|C_q^n\rangle $ is a superposition of $N(N-1)\cdots (N-n+1)/n!$
possible states $|b_1\cdots c_{i_1}\cdots c_{i_n}\cdots b_N\rangle $ in
which $n$ atoms are in the level $|c\rangle $ whereas the remaining $N-n$
atoms are in the level $|b\rangle $. For one of these states, each of the $%
N-n$ atoms being in the level $|b\rangle $ can be excited to the level $%
|a\rangle $. So the state $|A_q^1,C_q^n\rangle $ includes $N(N-1)\cdots
(N-n)/n!$ such possible states $|b_1\cdots a_l\cdots c_{i_1}\cdots
c_{i_n}\cdots b_N\rangle $ with equal possibility. The state $%
|A_q^1,C_q^n\rangle $ has been normalized. The phase factor related to the
excited atom $l$ being in level $|a\rangle $ is $\exp [i(k_s+q)z_l]$,
because the transition of the level $|a\rangle $ to the ground level $%
|b\rangle $ is connected with the signal field of the wavevector $k_s+q$.
State (\ref{C1}) can be obtained by the following operation 
\begin{eqnarray}
&&N\rho _{ac}(0)|C_q^n\rangle  \nonumber \\
&=&\left( \sum_{l=1}^N|a_l\rangle \langle c_l|\exp [ik_cz_l]\right)
|C_q^n\rangle  \tag{C2}  \label{C2} \\
&=&\sqrt{\frac{n!}{N\cdots (N-n+1)}}\sum_{l=1}^N\sum_{\{i_n\}}{}^{\prime
\prime }|a_l\rangle \langle c_l|b_1\cdots c_{i_1}\cdots c_{i_n}\cdots
b_N\rangle  \nonumber \\
&&\times \exp [i(k_s+q-k_c)(z_{i_1}+\cdots +z_{i_n})]\exp [ik_cz_l] 
\nonumber \\
&=&\sqrt{\frac{n!}{N\cdots (N-n+1)}}\sum_{l\neq
\{i_{n-1}\}}\sum_{\{i_{n-1}\}}{}^{\prime \prime }|b_1\cdots a_l\cdots
c_{i_1}\cdots c_{i_{n-1}}\cdots b_N\rangle  \nonumber \\
&&\times \exp [i(k_s+q-k_c)(z_{i_1}+\cdots +z_{i_{n-1}})]\exp [i(k_s+q)z_l] 
\nonumber \\
&=&\sqrt{\frac{n!}{N\cdots (N-n+1)}}\sqrt{\frac{N\cdots (N-n+1)}{(n-1)!}}%
|A_q^1,C_q^{n-1}\rangle  \nonumber  \label{C2} \\
&=&\sqrt{n}|A_q^1,C_q^{n-1}\rangle ,  \nonumber
\end{eqnarray}
where the atomic operator $\rho _{ac}(0)$ has been defined in Eq. (\ref{37b}%
). Similarly, we have 
\begin{eqnarray}
&&N\rho _{ab}(q)|C_q^n\rangle  \nonumber \\
&=&\left( \sum_{l=1}^N|a_l\rangle \langle b_l|\exp [i(k_s+q)z_l]\right)
|C_q^n\rangle  \tag{C3}  \label{C3} \\
&=&\sqrt{\frac{n!}{N\cdots (N-n+1)}}\sum_{l=1}^N\sum_{\{i_n\}}{}^{\prime
\prime }|a_l\rangle \langle b_l|b_1\cdots c_{i_1}\cdots c_{i_n}\cdots
b_N\rangle  \nonumber \\
&&\times \exp [i(k_s+q-k_c)(z_{i_1}+\cdots +z_{i_n})]\exp [i(k_s+q)z_l] 
\nonumber \\
&=&\sqrt{\frac{n!}{N\cdots (N-n+1)}}\sum_{l\neq
\{i_n\}}\sum_{\{i_n\}}{}^{\prime \prime }|b_1\cdots a_l\cdots c_{i_1}\cdots
c_{i_n}\cdots b_N\rangle  \nonumber \\
&&\times \exp [i(k_s+q-k_c)(z_{i_1}+\cdots +z_{i_n})]\exp [i(k_s+q)z_l] 
\nonumber \\
&=&\sqrt{\frac{n!}{N\cdots (N-n+1)}}\sqrt{\frac{N\cdots (N-n)}{n!}}%
|A_q^1,C_q^n\rangle  \nonumber  \label{C3} \\
&=&\sqrt{N-n}|A_q^1,C_q^n\rangle .  \nonumber
\end{eqnarray}

The two interactions induced by two fields in the interaction Hamiltonian
interfere destructively for the dark state. Using Eqs. (\ref{C2}) and (\ref
{C3}), for the exact expression of the dark state Eq. (\ref{44}), one
obtains 
\begin{eqnarray}
&&\Omega N\rho _{ac}(0)|D_q^n\rangle  \tag{C4}  \label{C4} \\
&=&\Omega N\rho _{ac}(0)\sum_{m=0}^n(-1)^m\sqrt{\frac{n(n-1)\cdots (n-m+1)}{%
m!}}  \nonumber \\
&&\times \sqrt{\frac{N(N-1)\cdots (N-m+1)}{N^m}}\frac{\Omega ^{n-m}(g\sqrt{N}%
)^m}{(\Omega ^2+g^2N)^{n/2}}|n-m\rangle |C_q^m\rangle  \nonumber \\
&=&\sum_{m=1}^n(-1)^m\sqrt{\frac{n(n-1)\cdots (n-m+1)}{(m-1)!}}\sqrt{\frac{%
N(N-1)\cdots (N-m+1)}{N^m}}  \nonumber  \label{72} \\
&&\times \frac{\Omega ^{n-m+1}(g\sqrt{N})^m}{(\Omega ^2+g^2N)^{n/2}}%
|n-m\rangle |A_q^1,C_q^{m-1}\rangle .  \nonumber
\end{eqnarray}
and 
\begin{eqnarray}
&&gNa(q)\rho _{ab}(q)|D_q^n\rangle  \tag{C5}  \label{C5} \\
&=&gNa(q)\rho _{ab}(q)\sum_{m=0}^n(-1)^m\sqrt{\frac{n(n-1)\cdots (n-m+1)}{m!}%
}  \nonumber  \label{71} \\
&&\times \sqrt{\frac{N(N-1)\cdots (N-m+1)}{N^m}}\frac{\Omega ^{n-m}(g\sqrt{N}%
)^m}{(\Omega ^2+g^2N)^{n/2}}|n-m\rangle |C_q^m\rangle  \nonumber \\
&=&\sum_{m=0}^{n-1}(-1)^m\sqrt{\frac{n(n-1)\cdots (n-m)}{m!}}\sqrt{\frac{%
N(N-1)\cdots (N-m)}{N^{m+1}}}  \nonumber \\
&&\times \frac{\Omega ^{n-m}(g\sqrt{N})^{m+1}}{(\Omega ^2+g^2N)^{n/2}}%
|n-m-1\rangle |A_q^1,C_q^m\rangle ,  \nonumber
\end{eqnarray}
If we set index $m\rightarrow m+1$ in Eq. (\ref{C4}), it is the exact same
as Eq. (\ref{C5}) but with an opposite sign. Therefore, one obtains 
\begin{equation}
\lbrack gNa(q)\rho _{ab}(q)+\Omega N\rho _{ac}(0)]|D_q^n\rangle =0.  \tag{C6}
\label{C6}
\end{equation}
Resulting in, for the exact resonant mode $q=0$, the dark states $%
|D_{q=0}^n\rangle $ being the eigenstates with the null eigenvalue of the
interaction Hamiltonian (\ref{38}). We note that Eqs. (\ref{C4}) -- (\ref{C6}%
) hold exactly for the exact expression of the dark state (\ref{44}). For
the approximate expression of the dark state (\ref{45}), Eqs. (\ref{C4}) -- (%
\ref{C6}) are also satisfied as long $N>>n$.

\end{document}